# The data center of tomorrow is made up of heterogeneous accelerators

Xavier Vasques

IBM Technology, France

## Abstract

The data center of tomorrow is a data center made up of heterogeneous systems, which will run heterogeneous workloads. The systems will be located as close as possible to the data. Heterogeneous systems will be equipped with binary, biological inspired and quantum accelerators. These architectures will be the foundations to address challenges. Like an orchestra conductor, the hybrid cloud will make it possible to set these systems to music thanks to a layer of security and intelligent automation.

# Introduction

Classical computing has experienced remarkable progress guided by Moore's Law. The law tells us that every two years, we double the number of transistors in a processor and at the same time we increase performance by two or reduce costs by two. This pace has slowed down over the past decade and we are now seeing a limit. This forces a transition. We must rethink information technology (IT) and in particular move towards heterogeneous system architectures with specific accelerators in order to meet the need for performance (1). The progress that has been made on raw computing power has nevertheless brought us to a point where biologically inspired computing models are now highly regarded in the state of the art (2) (3) (4).

Artificial intelligence (AI) is also an area bringing opportunities for progress but also challenges. The capabilities of AI have greatly increased in their ability to interpret and analyze data. AI is also demanding in terms of computing power because of the complexity of workflows. At the same time, AI can also be applied to the management and optimization of entire IT systems (1).

In parallel with conventional or biological-inspired accelerators, programmable quantum computing is emerging thanks to several decades of investment in research in order to overcome traditional physical limits. This new era of computing will potentially have the capacity to make calculations that are not possible today with conventional computers. Future systems will need to integrate quantum computing capabilities to perform specific calculations. Research is advancing rapidly. IBM made programmable quantum computers available to the cloud for the first time in May 2016 and announced its ambition to double the quantum volume each year named as Gambetta's law.

The cloud is also an element that brings considerable challenges and opportunities in IT and it has an important role to play. The data centers of tomorrow will be piloted by the cloud and equipped with heterogeneous systems that will run heterogeneous workloads in a secure manner. The data will no longer be centralized or decentralized but will be organized as hubs. The computing power will have to be found at the source of the data which today reaches extreme volumes. Storage systems are also full of challenges to improve availability, performance, management but also data fidelity. We have to design architectures allowing extracting more and more complex and often regulated data which poses multiple challenges, in particular security, encryption, confidentiality or traceability.

The future of computing, described by Dario Gil and William Green (1), will be built with heterogeneous systems made up of classical computing called binary or bit systems, biologically inspired computing and quantum computing called quantum or qubit systems (1). These heterogeneous components will be orchestrated and deployed by a hybrid cloud architecture that masks complexity while allowing the secure use and sharing of private and public systems and data.

## 1. Binary systems

The first binary computers were built in the 1940s: Colossus (1943) and then ENIAC (IBM - 1945). Colossus was designed to decrypt German secret messages and the ENIAC computer was designed to calculate ballistic trajectories. The ENIAC (Electronic Numerical Integrator and Computer), is in 1945 the first fully electronic built to be Turing-complete: it can be reprogrammed to solve, in principle, all the computational problems. The ENIAC was programmed by women, named "ENIAC women." The most famous of them were Kay McNulty, Betty Jennings, Betty Holberton, Marlyn Wescoff, Frances Bilas and Ruth Teitelbaum. These women had previously performed ballistic calculations on mechanical desktop computers for the military. The ENIAC then weighed 30 tons, occupied an area of 72 m2 and consumed 140 kilowatts.

Regardless of the task performed by a computer, the underlying process is always the same: an instance of the task is described by an algorithm which is translated into a sequence of 0 and 1, to give rise to execution in the processor, memory and input / output devices of the computer. This is the basis of binary calculation which in practice is based on electrical circuits provided with transistors which can be in two modes: " ON " allowing the current to pass and " OFF " the current does not pass. From these 0 and these 1 we have therefore developed over the past 80 years a classical information theory constructed from boolean operators (XAND, XOR), words (Bytes), and a simple arithmetic based on the following operations : " 0 + 0 = 0, 0 + 1 = 1 + 0 = 1, 1 + 1 = 0 (with restraint), and check if 1 = 1 , 0 = 0 and 1 ≠ 0. Of course, from these operations it is possible to build much more complex operations that computers can perform millions of billions of times per second for the most powerful of them. All this has become so "natural" that we completely forget that each transaction on a computer server, on a PC, a calculator, a smartphone breaks down into these basic binary operations. In a computer, these 0 and 1 are contained in "BInary digiTs " or " bits" which represent the smallest amount of information contained in a computer system. The electrical engineer and mathematician Claude Shannon (1916-2001) was one of the founding fathers of information theory. For 20 years, Claude Shannon worked at the Massachusetts Institute of Technology (MIT) and in addition to his academic activities, he worked at the Bell laboratories. In 1949, he married Madame Moore. During the Second World War, Claude Shannon worked for the American secret services, in cryptography, in order to locate messages hidden in German codes. One of his essential contributions concerns the theory of signal transmission (5) (6) (7). It is in this context that he developed an information theory, in particular by understanding that any data, even voice or images, can be transmitted using a sequence of 0 and 1.

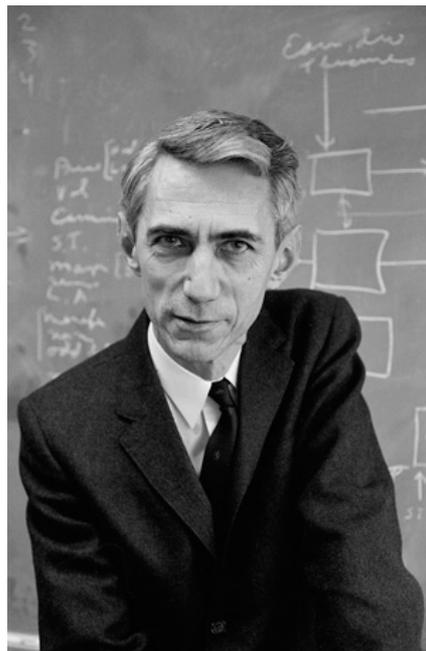

Photo: Alfred Eisenstaedt/The LIFE Picture Collection/Getty Images (40)

The binary used by conventional computers appeared in the middle of the 20th century, when mathematics and information were combined in a new way to form information theory, launching both the computer industry and telecommunications. The strength of the binary lies in its simplicity and reliability. A bit is either zero or one, a state which can be easily measured, calculated, communicated or stored (1). This powerful theory allowed to build today's systems that are running critical workloads around the world. Thanks to this method, various calculations and data storage systems have emerged up to the storage of digital data on a DNA molecule (8).

Today, we have examples of binary systems with incredible possibilities. A first example is the mainframe today called IBM Z. An IBM Z processor, Single Chip Module (SCM), uses Silicon On Insulator (SOI) technology at 14 nm. It contains 9.1 billion transistors. There are 12 cores per PU SCM at 5.2 GHz. This technology allows with a single system to be able to process 19 billion encrypted transactions per day and 1000 billion web transactions per day. All IBM Z installed in the world today process 87% of bank card transactions and 8 trillion payments per year (9).

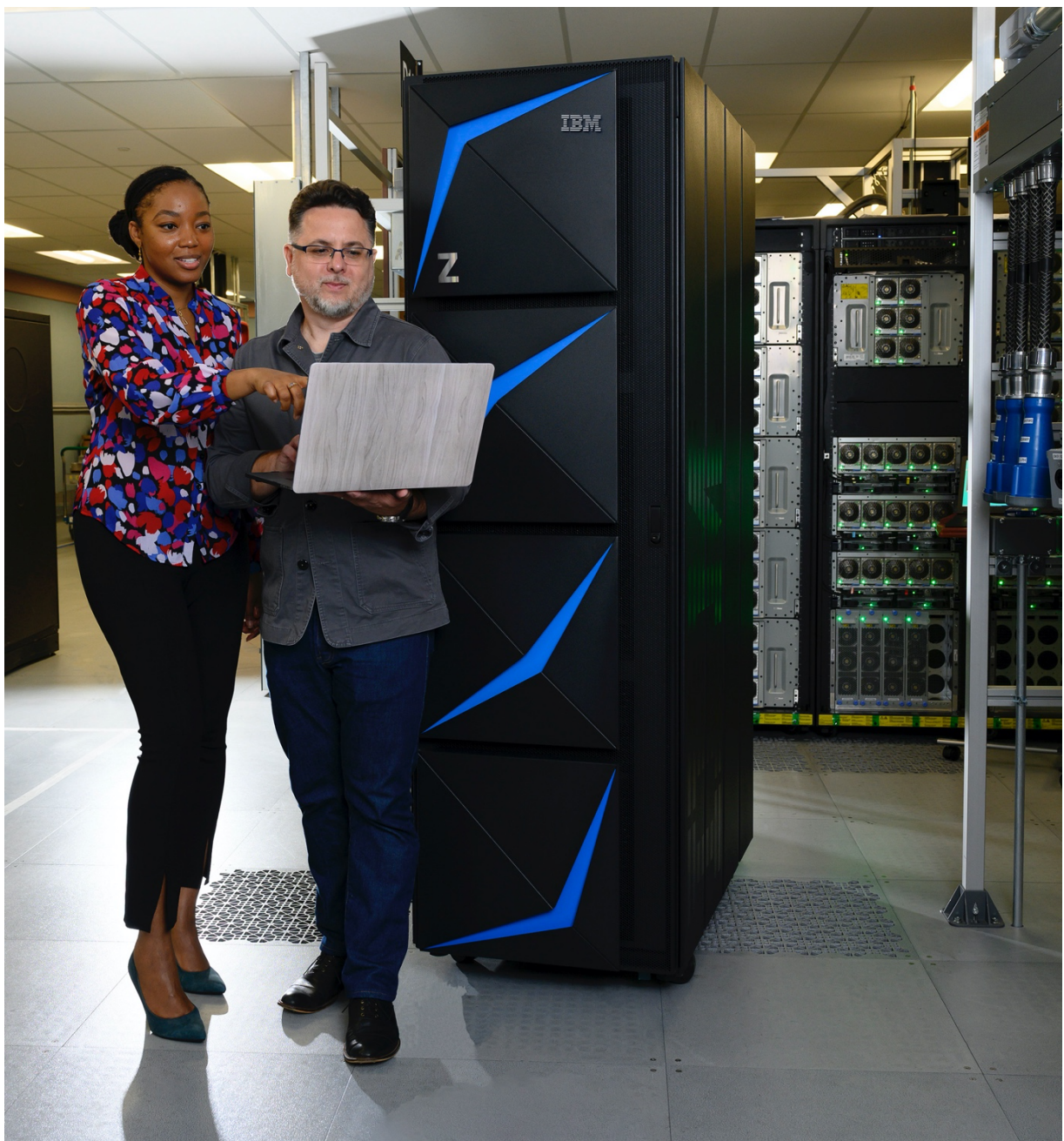

Source : IBM News Room (41)

We can also cite one of the most powerful computers in the world, Summit and Sierra respectively. They are located in the Oak Ridge Laboratory in Tennessee and the National Lawrence Laboratory in Livermore, California. These computers help model supernovas or new materials, explore solutions against cancer, study genetics and the environment. Summit is capable of delivering a computer power of 200 petaflops with a storage capacity of 250 petabytes. It is composed of 9216 IBM Power9 CPUs, 27648 NVIDIA Tesla GPUs and a network communication of 25 Gigabyte per second between the nodes. Despite everything, even the most powerful computer in the world, equipped with GPU accelerators, cannot calculate everything.

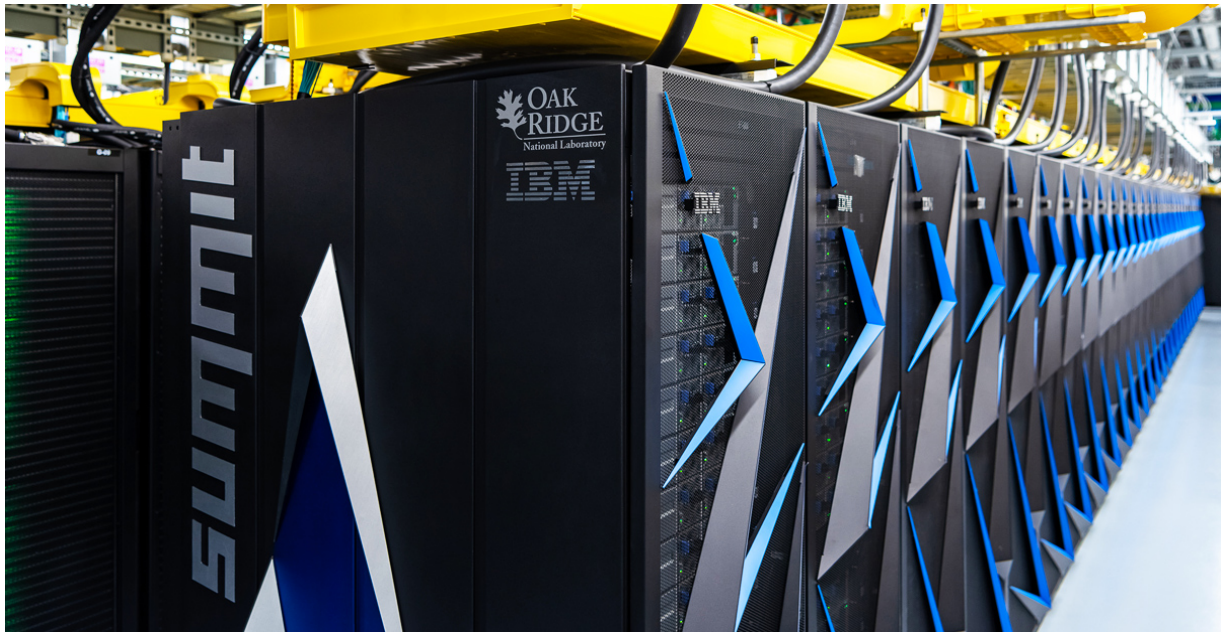
**Source : IBM News Room (42)**

Today, this type of technology is essential for medical research. And we saw it during the Covid-19 crisis. We can take the example of using the power of supercomputers with the HPC COVID-19 consortium (https://covid19-hpc-consortium.org). It is a public-private effort initiated by several institutions including IBM aimed at making the power of supercomputers available to researchers working on projects related to COVID-19 to help them identify potential short-term therapies for patients affected by the virus. Since its launch in March 2020, the Consortium's compute capacity has nearly doubled to 600 petaflops (one trillion floats per second), up from 330 petaflops in March.

Together, the Consortium has helped support many research projects, including understanding how long respiratory droplets persist in the air. This research by a team at Utah State University simulated the dynamics of indoor aerosols, providing insight into how long respiratory droplets linger in the air. They found that droplets from breathing linger in the air much longer than previously thought, due to their small size compared to droplets from coughs and sneezes. Another project concerns research into the reuse of drugs for potential treatments. A project by a Michigan State University team looked at data from about 1,600 FDA-approved drugs to see if there are possible combinations that could help treat COVID-19. They found at least two drugs approved by the FDA to be promising: proflavin, a disinfectant against many bacteria, and chloroxin, another antibacterial drug.

In France, a collaboration between the Institut Pasteur and IBM France is another example showing the need for acceleration (43). As we may have thousands of candidate molecules for potential therapeutic treatment, the use of accelerated systems and deep learning allows the best matches to be filtered in order to provide a selection of chemical compounds capable of attaching to pathogen proteins. By doing this, the drug design process could be sped up considerably. AI will also help researchers to better profile the protein-protein interactions involved in the development of pathologies, as well as to better understand the dynamics of infections in human cells. Thanks to this innovative approach, the development cycle of therapeutic treatment could be accelerated, potentially going from several years to a few months or even a few weeks, while saving millions of euros.

Computers, smartphones and their applications, the internet that we use in our everyday lives work with 0's and 1's. The binary coupled with Moore's law, a 50-year heritage, has made it possible to build systems robust and reliable. For 50 years, we have seen incremental, linear evolution to have performance gains. The next few years will bring us their lots of innovations in order to have performance gains, particularly in terms of materials, control processes or etching method: we are talking about three-dimensional transistors, extreme ultraviolet lithography or new materials such as hafnium. or germanium. The binary therefore continues to evolve and will play a central role in the data center of tomorrow.

Recently, IBM took a big step forward in chip technology by manufacturing the first 2 nm chip to squeeze 50 billion transistors onto a fingernail-sized chip. The architecture can help processor manufacturers improve performance by 45% with the same amount of power as today's 7 nm chips, or the same level of performance using 75% less power. Mobile devices with 2nm-based processors could have up to four times the battery life of those with 7 nm chipsets. Laptops would benefit from an increase in the speed of these processors, while autonomous vehicles will detect and respond to objects faster. This technology will benefit data center energy efficiency, space exploration, artificial intelligence, 5G and 6G and quantum computing.

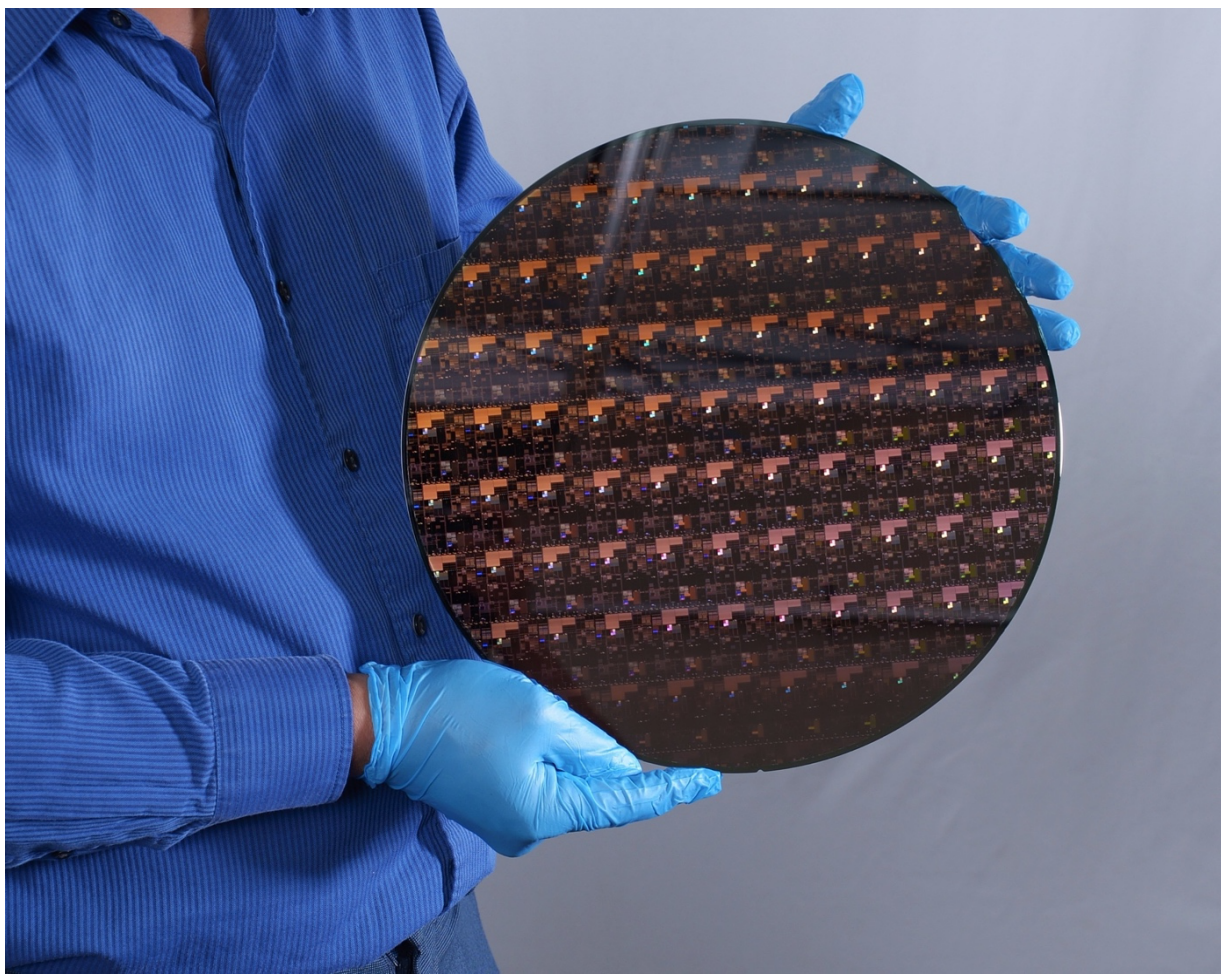

Despite comments about the limitations of Moore's Law and the bottlenecks of current architectures, innovations continue and binary will play a central role in the data center of tomorrow. Nevertheless, some challenges cannot be addressed only with binary. For these challenges, we need to rethink computing with new approaches drawing inspiration from nature such as biology and physics.

## 2. Biologically inspired systems

DNA is also a source of inspiration. A team recently published in the journal Science the capacity to store digital data on a DNA molecule. They were able to store an operating system, a French film from 1895 (L'Arrivée d'un train à La Ciotat by Louis Lumière), a scientific article, a photo, a virus and a $ 50 gift card in DNA strands and retrieve the data without errors.
Indeed, a DNA molecule is intended to store information by nature. Genetic information is four nitrogenous bases that make up a DNA molecule (A, C, T, and G). Today it is possible to transcribe digital data into a new code. DNA sequencing then makes it possible to read the stored information. Encoding is automated through software. A DNA molecule is 3 billion nucleotides (nitrogenous base). In one gram of DNA, 215 petabytes of data can be stored. It would be possible to store all the data created by humans in one room. In addition, DNA can theoretically keep data in perfect condition for an extremely long time. Under ideal conditions, it is estimated that DNA could still be deciphered after several million years thanks to "longevity genes". DNA can withstand the most extreme weather conditions. The main weak points today are high cost and processing times which can be extremely long.

The term AI as such appeared in 1956. Several American researchers, including John McCarthy, Marvin Minsky, Claude Shannon and Nathan Rochester of IBM, very advanced in research that used computers for other than scientific calculations, met at the University of Dartmouth, in the United States. Three years after the Dartmouth seminar, the two fathers of AI, McCarthy and Minsky, found the AI lab at MIT. There was a lot of investment, too much ambition, to imitate the human brain, and a lot of hope that was not realized at the time. The promises were broken. A more pragmatic approach appeared in the 1970s and 1980s, which saw the emergence of machine learning and the reappearance of neural networks in the late 1980s. This more pragmatic approach, the increase in computing power and the explosion of data has made it possible for AI to be present in all areas today, it is a transversal subject. The massive use of AI poses some challenges such as the need to label the data at our disposal. The problem with automation is that it requires a lot of manual work. AI needs education. This is done by tens of thousands of workers around the world which does not really look like what you might call a futuristic vision. Another challenge is the need for computing power. AI needs to be trained and for this AI is more and more greedy in terms of calculations. The training requires a doubling of the computing capacities every 3.5 months (10).

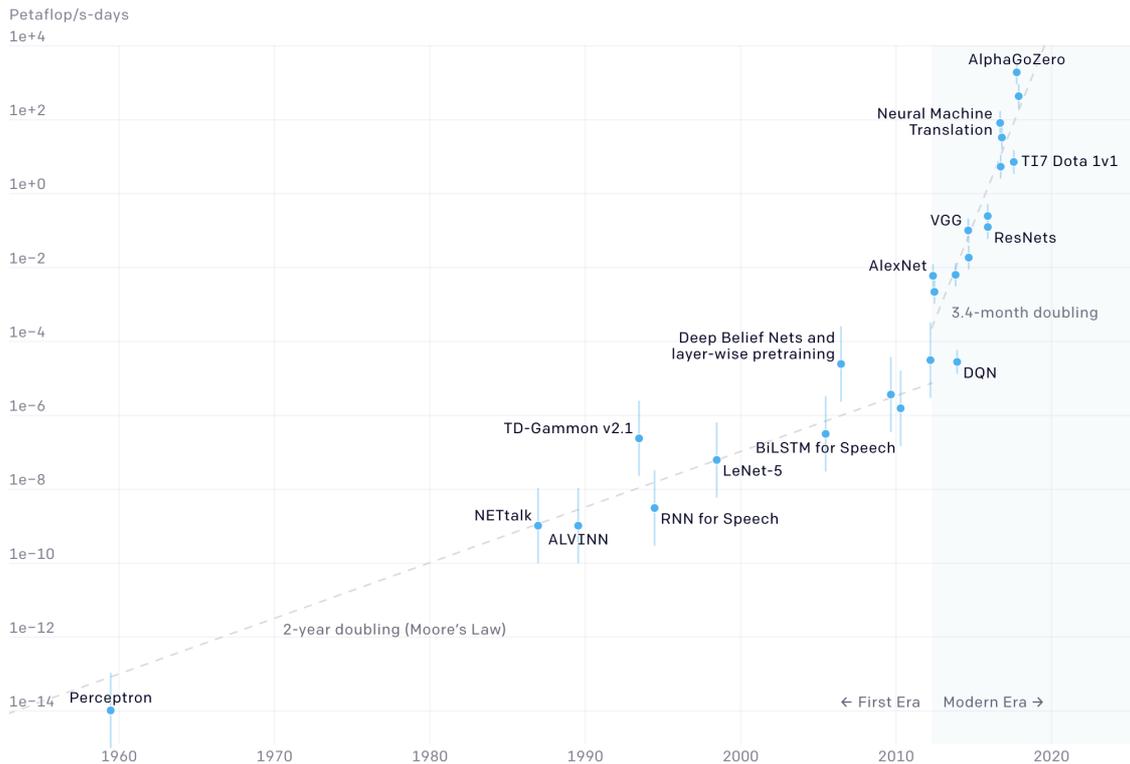

Source: AI and Compute, OpenAI, https://openai.com/blog/ai-and-compute/#fn1

Several approaches are currently used and envisaged. Today for example, as for the Summit supercomputer, the calculation of certain workloads is deported to accelerators such as GPUs. There are others such as FPGAs (Field Programmable Gate Arrays or "programmable logic networks") which can realize the desired digital functions. The advantage is that the same chip can be used in many different electronic systems.

Progress in the field of neuroscience will allow the design of processors directly inspired by the brain. The way our brain transmits information is not binary. And it is thanks to Santiago Ramón y Cajal (1852-1934), Spanish histologist and neuroscientist, Nobel Prize in physiology or medicine in 1906 with Camillo Golgi that we know better the architecture of the nervous system. Neurons are cellular entities separated by fine spaces, synapses, and not fibers of an unbroken network (11). The axon of a neuron transmits nerve impulses, action potential, to target cells. The next step in developing new types of AI-inspired and brain-inspired processors is to think differently about how we compute today. Today one of the major performance problems is the movement of data between the different components of the von Neumann architecture: processor, memory, storage. It is therefore imperative to add analog accelerators. What dominates numerical calculations today and in particular deep learning calculations is floating point multiplication. One of the methods envisaged as an effective means of gaining computational power is to go back by reducing the precision also called approximate calculation. For example, 16-bit precision engines are more than 4x smaller than 32-bit precision engines (1). This gain increases performance and energy efficiency. In simple terms, in approximate calculation, we can make a compromise by exchanging numerical precision for the efficiency of calculation. Certain conditions are nevertheless necessary such as developing algorithmic improvements in parallel to guarantee iso-precision (1). IBM recently demonstrated the success of this approach with 8-bit floating-point numbers, using new techniques to maintain the accuracy of gradient calculations and updating weights during backpropagation (12) (13). Likewise, for the inference of a model resulting from the deep learning algorithm training, the unique use of whole arithmetic on 4 or 2 precision bits achieves accuracy comparable to a range of popular models and data sets (14). This progression will lead to a dramatic increase in computing capacity for deep learning algorithms over the next decade.

Analog accelerators are another way of avoiding the bottleneck of von Neumann's architecture (15) (16). The analog approach uses non-volatile programmable resistive processing units (RPUs) that can encode the weights of a neural network. Calculations such as matrix or vector multiplication or the operations of matrix elements can be performed in parallel and in constant time, without movement of the weights (1). However, unlike digital solutions, analog AI will be more sensitive to the properties of materials and intrinsically sensitive to noise and variability. These factors must be addressed by architectural solutions, new circuits and algorithms. For example, analogous non-volatile memories (NVMs) (17) can effectively speed up backpropagation algorithms. By combining long-term storage in phase-change memory (PCM) devices, quasi-linear updating of conventional CMOS capacitors and new techniques to eliminate device-to-device variability, significant results began to emerge for the calculation of Deep Neural Network (18) (19) (20). The research also embarked on a quest to build a chip directly inspired by the brain (21). In an article published in Science (22), IBM and its university partners have developed a processor called SyNAPSE which is made up of a million neurons. The chip consumes only 70 milliwatts and is capable of 46 billion synaptic operations per second, per watt, literally a synaptic supercomputer holding in a hand. We have moved from neuroscience to supercomputers, a new computing architecture, a new programming language, algorithms, applications and now a new chip called TrueNorth (23). TrueNorth is a neuromorphic CMOS integrated circuit produced by IBM in 2014. It is a many-core processor network, with 4096 cores, each having 256 programmable simulated neurons for a total of just over one million neurons. In turn, each neuron has 256 programmable synapses allowing the transport of signals. Therefore, the total number of programmable synapses is slightly more than 268 million. The number of basic transistors is 5.4 billion. Since memory, computation and communication are managed in each of the 4096 neurosynaptic cores, TrueNorth bypasses the bottleneck of the von Neumann architecture and is very energy efficient. It has a power density of 1 / 10,000 of conventional microprocessors.

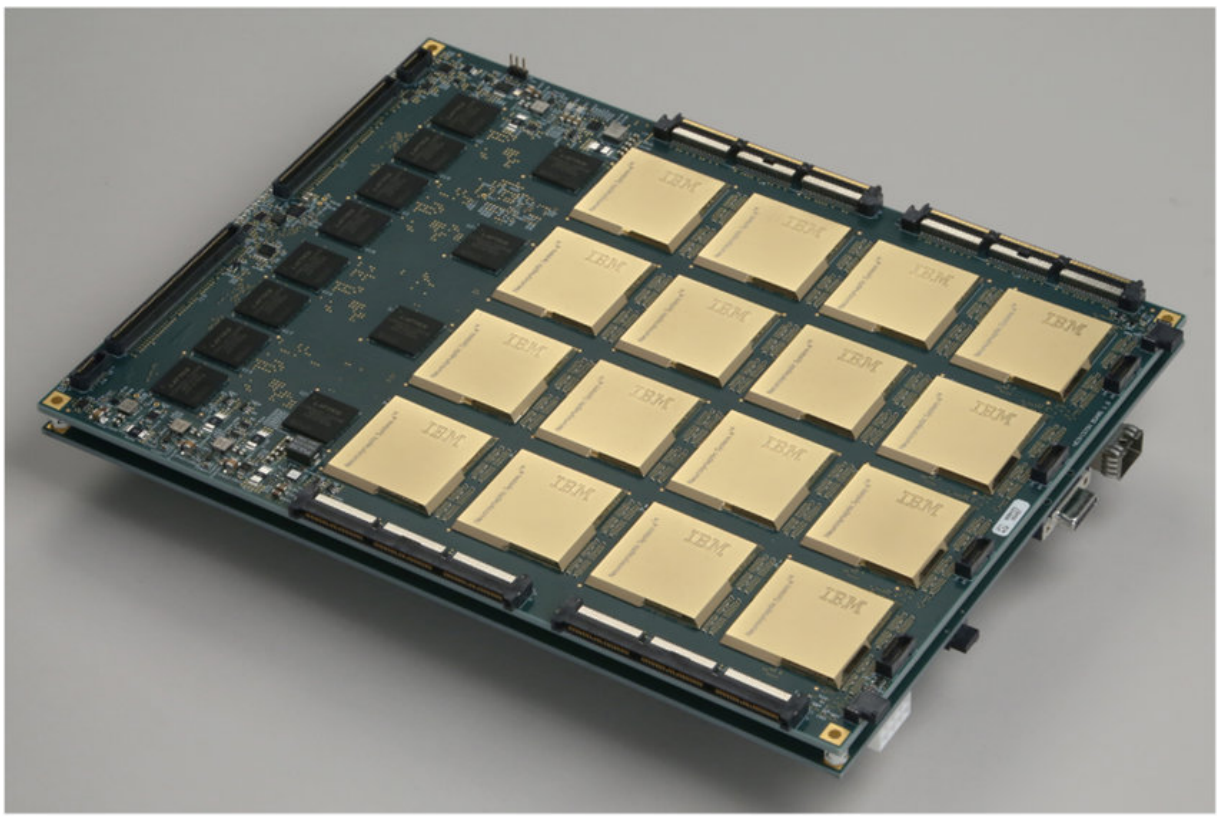

Source : https://www.research.ibm.com/articles/brain-chip.shtml

## 3. Quantum systems: Qubits

In an article published in Nature, IBM physicists and engineers have described how they achieved the feat of writing and reading data in a Holmium atom, a rare-earth element. This is a symbolic step forward but proves that this approach works. That we might one day have atomic data storage. To compare what it means, imagine that we can store the entire iTunes library of 35 million songs on a device the size of a credit card. In the paper, the nanoscientists demonstrated the ability to read and write one bit of data on one atom. For comparison, today's hard disk drives use 100,000 to one million atoms to store a single bit of information.

Of course, we cannot avoid discussing quantum computing. Quantum bits - or qubits - combine physics with information and are the basic units of a quantum computer. Quantum computers use qubits in a computational model based on the laws of quantum physics. Properly designed quantum algorithms will be capable of solving problems of great complexity by exploiting quantum superposition and entanglement to access an exponential state space, then by amplifying the probability of calculating the correct response by constructive interference. It was from the beginning of the 1980s, under the impulse of the physicist and Nobel Prize winner Richard Feynman, that the idea of the design and development of quantum computers was born: Where a " classical " computer works with bits of values 0 or 1, the quantum computer uses the fundamental properties of quantum physics and is based on "quantum bits (qubits)". Beyond this technological progress, quantum computing opens the way to the processing of computer tasks whose complexity is beyond the reach of our current computers. But let's start from the beginning.

At the beginning of the 20th century, the theories of so-called classical physics were unable to explain certain problems observed by physicists. They must therefore be reformulated and enriched. Under the impetus of scientists, it will evolve initially towards a "new mechanics" which will become "wave mechanics" and finally "quantum mechanics." Quantum mechanics is the mathematical and physical theory that describes the fundamental structure of matter and the evolution over time and space of the phenomena of the infinitely small. An essential notion of quantum mechanics is the duality "wave - particle." Until the 1890s, physicists considered that the world was composed of two types of objects or particles: on the one hand those which have a mass (like electrons, protons, neutrons, atoms…), and, on the other hand, those who do not have one (like photons, waves…). For the physicists of the time, these particles are governed by the laws of Newtonian mechanics for those which have a mass and by the laws of electromagnetism for the waves. We therefore had two theories of " Physics " to describe two different types of objects. Quantum mechanics invalidate this dichotomy and introduce the fundamental idea of the particle wave duality. The particles of matter or the waves must be treated with the same laws of physics. It is the advent of wave mechanics which will become quantum mechanics a few years later. Big names are associated with the development of quantum mechanics such as Niels Bohr, Paul Dirac, Albert Einstein, Werner Heisenberg, Max Planck, Erwin Schrödinger and many others. Max Planck and Albert Einstein, being interested in the radiation emitted by a heated body and in the photoelectric effect, were the first to understand that the exchanges of light energy could only be done by "packet." Moreover, Albert Einstein obtained the Nobel Prize in physics following the publication of his theory on the quantified aspect of energy exchanges in 1921. Niels Bohr extended the quantum postulates of Planck and Einstein from light to matter, by proposing a model reproducing the spectrum of the hydrogen atom. He obtained the Nobel Prize in physics in 1922, by defining a model of the atom which dictates the behavior of quanta of light. Passing from one energy level to another lower, the electron exchanges a quantum of energy. Step by step, rules were found to calculate the properties of atoms, molecules and their interactions with light.
From 1925 to 1927, a whole series of works by several physicists and mathematicians gave substance to two general theories applicable to these problems:
- The wave mechanics of Louis de Broglie and especially of Erwin Schrödinger;

- The matrix mechanics of Werner Heisenberg, Max Born and Pascual Jordan.

These two mechanics were unified by Erwin Schrödinger from the physical point of view, and by John von Neumann from the mathematical point of view. Finally, Paul Dirac formulated the synthesis or rather the complete generalization of these two mechanics, which today we call quantum mechanics. The fundamental equation of quantum mechanics is the Schrödinger equation.

Quantum computing started with a small, now famous, conference in 1981, jointly organized by IBM and MIT, on the physics of computing. The physicist Nobel laureate Richard Feynman challenged computer scientists to invent a new type of computer based on quantum principles to better simulate and predict the behavior of the actual material (24): "I'm not satisfied with all the analyzes that go with just classical theory, because nature is not classic, damn it, and if you want to make a simulation of nature, you'd better do it with quantum mechanics…".

Matter, Feynman explained, is made up of particles such as electrons and protons that obey the same quantum laws that would govern the operation of this new computer. Since then, scientists have tackled Feynman's double challenge: understanding the capabilities of a quantum computer and figuring out how to build one. Quantum computers will be very different from today's computers, not only in what they look like and how they are made, but, more importantly, in what they can do. We can also quote a famous phrase from Rolf Landauer, a physicist who worked at IBM: " Information is physical ." The computers are, of course, physical machines. It is therefore necessary to take into account the energy costs generated by calculations, recording and reading of the information bits as well as energy dissipation in the form of heat . In a context where the links between thermodynamics and information were the subject of many questions, Rolf Landauer , sought to determine the minimum amount of energy necessary to manipulate a single bit of information in a given physical system. There would therefore be a limit, today called the Landauer limit and discovered in 1961, which defines that any computer system is obliged to dissipate a minimum of heat and therefore consume a minimum of electricity. This research is fundamental because it shows that any computer system has a minimum thermal and electrical threshold that we cannot exceed. This will mean that we will reach the minimum consumption of a computer chip and that it will not be able to release less energy. It is not for now, but scientists explain that this limit will be especially present when designing quantum chips. Recent work by Charles Henry Bennett at IBM has consisted of a re-examination of the physical basis of information and the application of quantum physics to the problems of information flows. His work has played a major role in the development of an interconnection between physics and information.

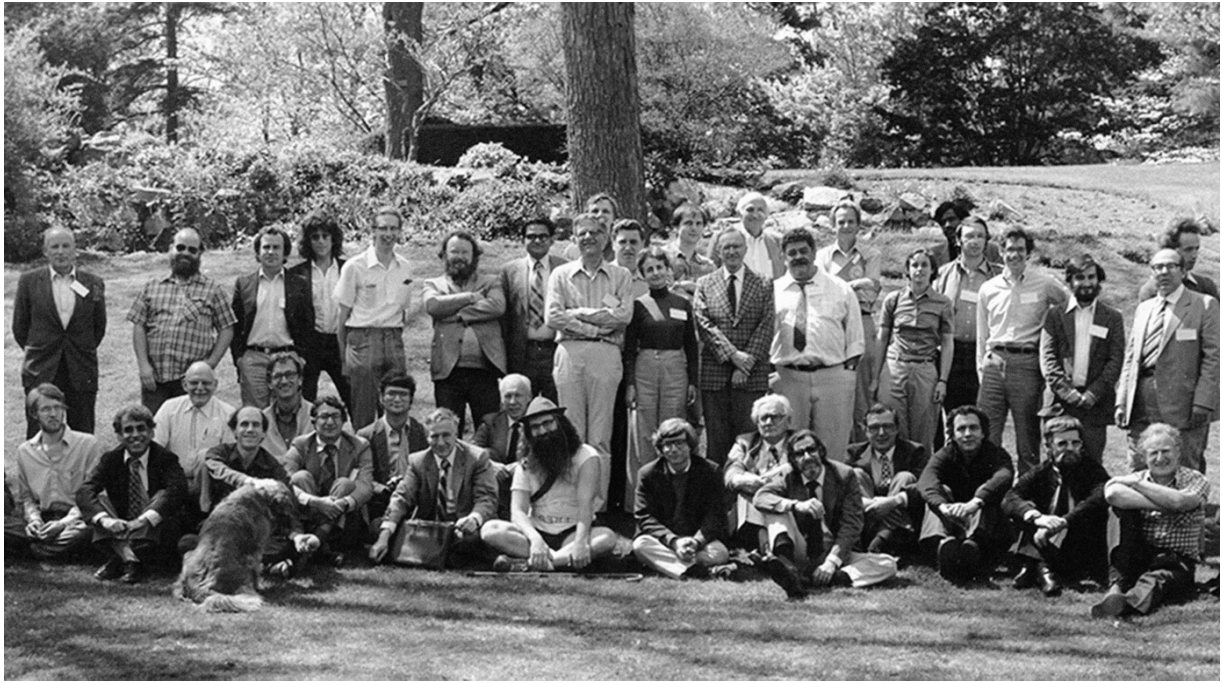
Source : https://research.ibm.com/blog/qc40-physics-computation

For a quantum computer the qubit is the basic entity, representing, like the "bit," the smallest entity allowing manipulating information. It has two fundamental properties of quantum mechanics: superposition and entanglement.

A quantum object (on a microscopic scale) can exist in an infinity of states (as long as one does not measure this state). A qubit can therefore exist in any state between 0 and 1. The qubits can take both the value 0 and the value 1, or rather "a certain amount of 0 and a certain amount of 1", as a combination linear of two states denoted $|0\rangle$ and $|1\rangle$, with the coefficients $\alpha$ and $\beta$. So, where a classic bit describes "only" 2 states (0 or 1), the qubit can represent an "infinity." This is one of the potential advantages of quantum computing from the point of view of information theory. We can get an idea of the superposition of states using the analogy of the lottery ticket: a lottery ticket is either winning or losing once we know the outcome of the game. However, before the draw, this ticket was neither a winner nor a loser. It just had a certain probability of being a winner and a certain probability of being a loser; it was both a winner and a loser at the same time. In the quantum world, all the characteristics of particles can be subject to this indeterminacy: for example, the position of a particle is uncertain. Before the measurement, the particle is neither at point A, nor at point B. It has a certain probability of being at point A and a certain probability of being at point B. However, after the measurement, the state of the particle is well defined: it is at point A or at point B.

Another amazing property of quantum physics is entanglement. When we consider a system composed of several qubits, they can sometimes "link their destiny," that is to say not to be independent of each other even if they are separated in the space (while the "classic" bits are completely independent of each other). This is called quantum entanglement. If we consider a system of two entangled qubits then the measurement of the state of one of these two qubits gives us an immediate indication of the result of an observation on the other qubit.

To naively illustrate this property, we can also use an analogy here: imagine two light bulbs, each in two different houses. By entangling them, it becomes possible to know the state of a bulb (on or off) by simply observing the second, because the two would be linked, entangled. And this, immediately and even if the houses are very far from each other.

This entanglement phenomenon makes it possible to describe correlations between qubits. If we increase the number of qubits, the number of these correlations increases exponentially: for N qubits there are 2n correlations. It is this property which gives the quantum computer the possibility of carrying out manipulations on gigantic quantities of values, quantities beyond the reach of a conventional computer.

The uncertainty principle discovered by Werner Heisenberg in 1927, tells us that, whatever our measuring tools, we are unable to precisely determine both the position and the speed of a quantum object (at the atomic scale). Either we know exactly where the object is and the speed will seem to fluctuate and become blurry, or we have a clear idea of the speed but its position will escape us.

Julien Bobroff in his book La quantique autrement : garanti sans équation ! describes the quantum experiment in three acts:

The first moment is before the quantum object behaves like a wave. The Schrödinger equation allows us to accurately predict how the latter spreads, how fast, in what direction, whether it spreads or contracts. Then it is decoherence that makes its appearance. Decoherence happens extremely quickly, almost instantaneously. It is at this precise moment that the wave comes into contact with a measuring tool (eg a fluorescent screen). This wave is forced to interact with the particles that make up this device. This is the moment when the wave is shrinking. The last step is the random choice among all the possible states. The draw is related to the shape of the wave function at the time of measurement. In fact, only the shape of the wave function at the end of the first act dictates how likely it is to appear here or there.

Another phenomenon of quantum mechanics is the tunnel effect. Julien Bobroff gives the example of a tennis ball. At the opposite if the ball, a quantum wave function only partially bounces against a barrier. A small part can tunnel through to the other side. This implies that if the particle is measured, it will sometimes materialize on the left of the wall, sometimes on the right.

A quantum computer therefore uses the laws of quantum mechanics to make calculations. It has to be under certain conditions, sometimes extreme, such as immersing a system in liquid helium to reach temperatures close to absolute zero, i.e. -273.15 ° C.

Building a quantum computer relies on the ability to develop a computer chip on which qubits are engraved. From a technological point of view, there are several ways of constituting qubits, they can be made of atoms, photons, electrons, molecules or superconductive metals. In most cases, in order to function, a quantum computer needs extreme conditions to operate, such as temperatures close to absolute zero. The choice of IBM for example, is to use superconducting qubits, constructed with aluminum oxides (this technology is also called: transmons qubits). As mentioned above to allow and guarantee the quantum effects (superposition and entanglement) the qubits must be cooled to a temperature as close as possible to absolute zero (i.e. around -273 ° C). At IBM this operating threshold is around 20 milliKelvin! IBM demonstrated the ability to design a single qubit in 2007 and 2016 and announced the availability in the cloud of a first operational physical system with 5 qubits and a development environment "QISKit" (Quantum Information Science Kit), allowing to design, test and optimize algorithms for commercial and scientific applications. The "IBM Q Experience" initiative is a first in the industrial world.

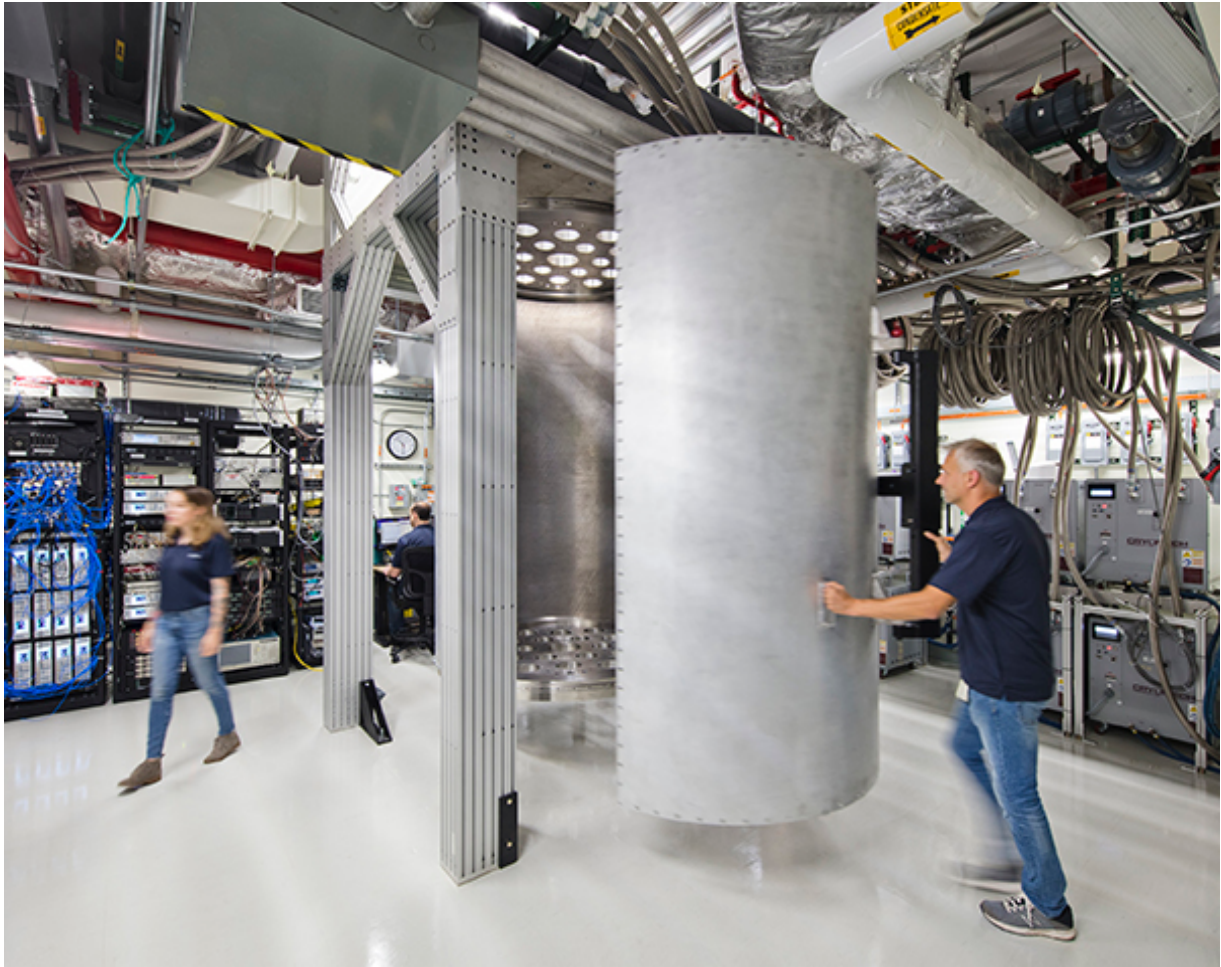
Source : https://www.ibm.com/blogs/research/2020/09/ibm-quantum-roadmap/

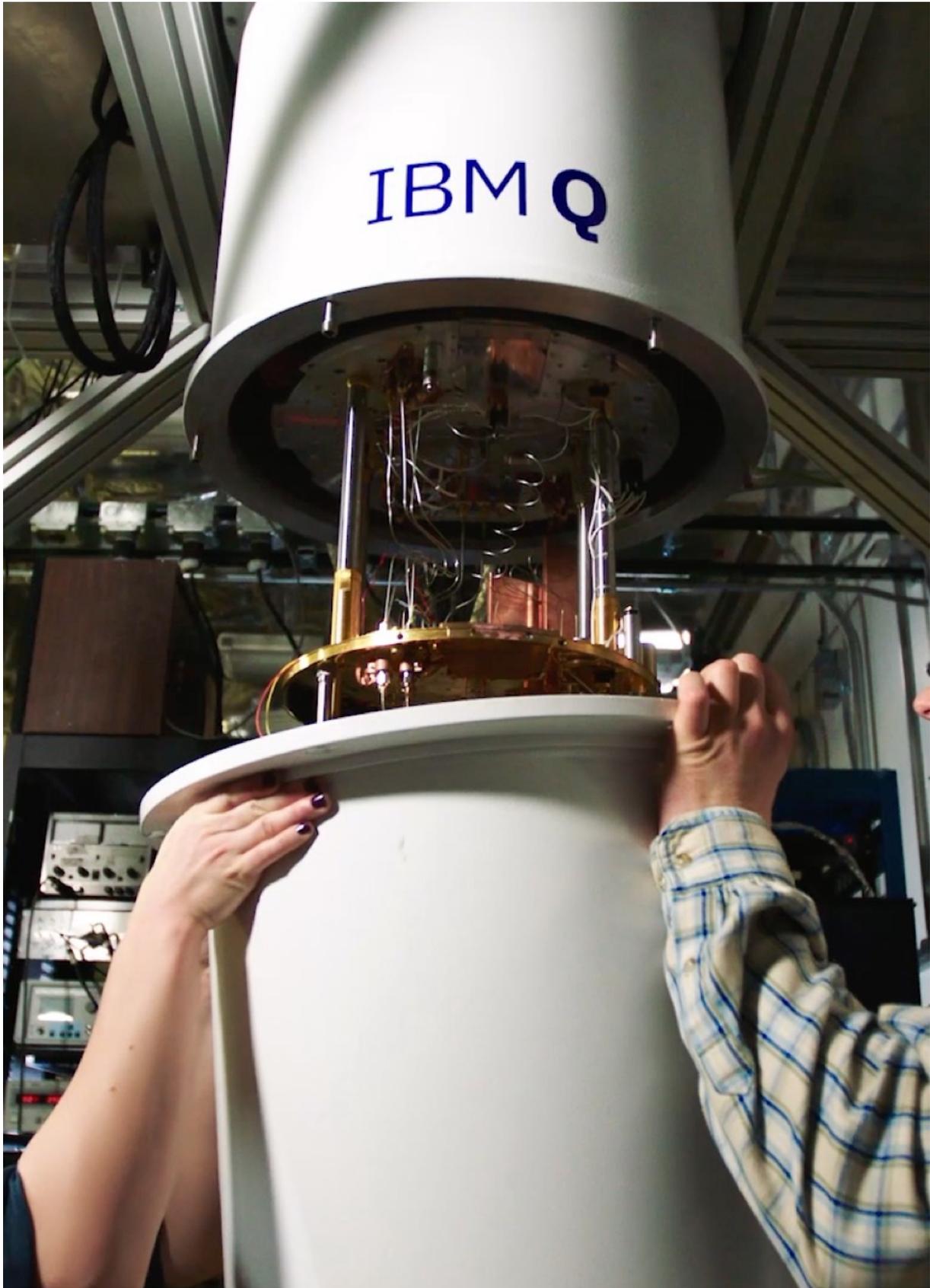

Source : https://www.ibm.com/quantum-computing/ibm-q-network/

IBM now has 32 quantum computers, including a 65-qubit system, and recently published its roadmap.

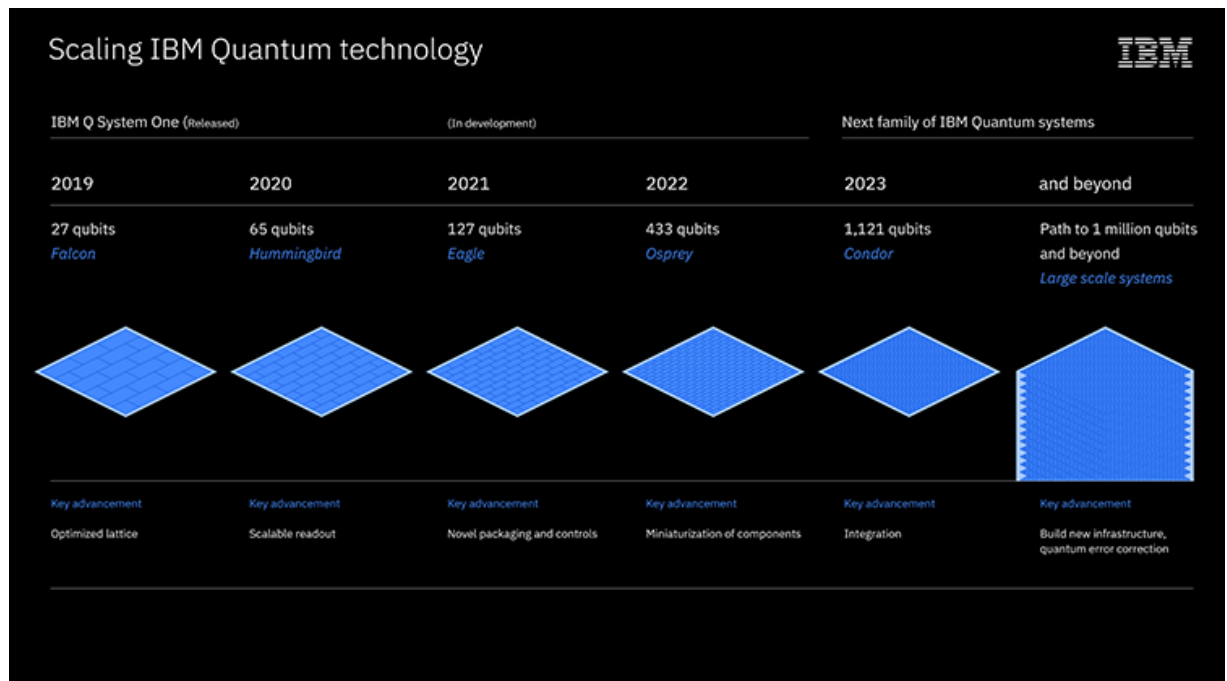

Source : https://www.ibm.com/blogs/research/2020/09/ibm-quantum-roadmap/

This openness to the public has enabled more than 300,000 users to initiate hundreds of billions of circuit runs on real hardware and simulators. Access to its machines has led to the publication of more than 400 research papers by non-IBM researchers. IBM has also built a network of over 140 members, known as the IBM Quantum Network, with privileged access to the latest quantum technologies to work in particular on use cases. IBM Quantum Network is made up of large groups, startups, universities and laboratories all over the world.

The number of qubits will progressively increase but this is not enough. In the race to develop quantum computers, beyond qubits, other components are essential. We speak of "quantum volume" as a relevant measure of performance and technological progress (26). Other measures are also offered by companies and laboratories. We also define "quantum advantage" is the point from which quantum computing applications will offer a significant and demonstrable practical advantage that exceeds the capabilities of conventional computers alone. The concept of quantum volume was introduced by IBM in 2017. It is beginning to spread to other manufacturers. Quantum volume is a measure that determines the power of a quantum computer system, taking into account both gate and measurement errors, crosstalk of the device, as well as the connectivity of the device and the efficiency of the circuit compiler. The quantum volume can be used for any NISQ quantum computer system based on gates and circuits. For example, if you lower the error rate of x10 without adding extra qubits you can have a quantum volume increase of 500x. On the contrary, if you add 50 additional qubits but you not decrease error rates, you can have an increase quantum volume of 0x. Adding qubits isn't everything.

Today's challenge researchers face are technological, such as stability over time. When you run a quantum algorithm on a real quantum computer, there are a lot of externalities that can disrupt the quantum state of your program, which is already fragile. Another technological challenge concerns the quantity of qubits that we will be able to take into account. Every time you increase the capacity of a quantum computer by one qubit, you reduce its stability. Another challenge is that we are going to be forced to rethink the entirety of the algorithm we know to adapt it to quantum computing.

And of course, you need to be able to run tasks on these machines, which is why IBM has developed a specific programming library called QISKit (Quantum Information Science Kit). It is an open-source library for the Python language, available on qiskit.org. Its development is very active, all the contributors, including IBM, regularly update the functionality of this programming environment.

Quantum computers will be added to conventional computers to address problems that are today unsolved. Conventional computers can, for example, calculate complex problems that a quantum computer cannot. There are problems that both classical and quantum computers can solve. And finally, challenges that a conventional computer cannot solve but that a quantum computer can address. Many applications are possible in the field of chemistry, materials science, machine learning or optimization.

For example, it is difficult for a classical computer to calculate exactly (that is to say without any approximation) the energy of the caffeine molecule, yet of average size with 24 atoms, it is a very complex problem (27) (28). We would need approximately $10^{48}$ bits to represent the energy configuration of a single caffeine molecule at a time t. That is almost the number of atoms admitted to earth which is $10^{50}$. But we believe it is possible to do this with 160 qubits. Today quantum computers are used to treat simple chemistry problems, that is to say with a small number of atoms, but the objective is, of course, to be able to address much more complex molecules (27) (28).

But that's not the only limitation. To give you a simple illustration, we can talk about the so-called itinerant seller problem, or more exactly today, the problem of the delivery of delivery trucks. If you want to choose the most efficient route for a truck to deliver packages to five addresses, there are 12 possible routes, so it is at least possible to identify the best one. However, as you add more addresses, the problem becomes exponentially more difficult - by the time you have 15 deliveries, there are over 43 billion possible routes, making it virtually impossible to find the best. For example, in 71 cities, the number of candidate paths is greater than $5 \times 10^{80}$.

Currently, quantum computing is suitable for certain algorithms such as optimization, machine learning or simulation. With this type of algorithms, the use cases apply in several industrial sectors. Financial services such as portfolio risk optimization, fraud detection, health (drug research, protein study, etc.), supply chains and logistics, chemicals, research for new materials or oil are all areas that will be primarily impacted.

We can address more specifically the future of medical research with quantum, which should eventually allow the synthesis of new therapeutic molecules. If we are to meet the challenge of climate change for example, we need to solve many problems such as designing better batteries, finding less energy-intensive ways to grow our food and planning our supply chains to minimize transport. Solving these problems effectively requires radically improved computational approaches in areas such as chemistry and materials science, as well as optimization and simulation - areas where classical computing face serious limitations.

For a conventional computer, making the product of two numbers and obtaining the result is a very simple operation: 7 * 3 = 21 or 6739 * 892721 = 6016046819. This remains true for very large numbers. But the opposite problem is much more complex. Knowing a large number N, it is more complicated to find P and Q such that: P x Q = N. It is this difficulty which is on the basis of current cryptographic techniques. For such a problem, by way of example, it is estimated that a problem which would last 1025 days on a conventional computer, could be resolved in a few tens of seconds on a quantum machine. We speak in this case of exponential acceleration. With quantum computers, we can approach problems in a whole new way by taking advantage of entanglement, superposition and interference: modeling the physical procedures of nature, performing a lot more scenario simulations, or getting better optimization solutions, find better models in AI / ML processes. In these categories of problem eligible for quantum computers there are many cases of optimization, in the fields of logistics (shortest path), finance (risk assessment, evaluation of asset portfolios), marketing, industry and the design of complex systems (29) (30) (31) (32) (33) (34) . The field of AI (35) (36) is also an active research field, and learning methods for artificial neural networks are starting to emerge, so it is the whole of the human activities concerned by the processing of information that is potentially concerned with the future of quantum computing. The domain of cybersecurity and cryptography is also a subject of attention. The Shor algorithm was demonstrated over 20 years ago and it could weaken the encryption commonly used on the internet. We will have to wait until the quantum machines are powerful enough to process this particular type of calculation, and, on the other hand, encryption solutions are already known and demonstrated beyond the reach of this algorithm. Quantum technology itself will also provide solutions to protect data.

Therefore, the field of quantum technologies and quantum computing in particular is considered strategic, and Europe, France and many other countries support research efforts in this area.

If we take the use cases by the industrial field, we can find many in banks and financial institutions: improve trading strategies, improve management of client portfolios and better analyze financial risks. A quantum algorithm in development, for example, could potentially provide quadratic acceleration when using derivative pricing - a complex financial instrument that requires 10,000 simulations to be valued on a conventional computer, would only require 100 quantum operations on a quantum device.

One of the use cases of quantum is the optimization of trading. It will also be possible for banks to accelerate portfolio optimizations such as Monte Carlo simulations. The simulation of buying and selling of products (trading) such as derivatives can be improved thanks to quantum computing. The complexity of trading activities in the financial markets is skyrocketing. Investment managers struggle to integrate real constraints, such as market volatility and changes in client life events, into portfolio optimization. Currently, the rebalancing of investment portfolios that follow market movements is strongly impacted by calculation constraints and transaction costs. Quantum technology could help reduce the complexity of today's business environments. The combinatorial optimization capabilities of quantum computing can enable investment managers to improve portfolio diversification, rebalance portfolio investments to more precisely respond to market conditions and investor objectives, and to streamline more cost-effective transaction settlement processes. Machine learning is also used for a portfolio optimization and scenario simulation. Banks and financial institutions such as hedge funds are increasingly interested because they see it as a way to minimize risks while maximizing gains with dynamic products that adapt according to new simulated data. Personalized finance is also an area explored. Customers demand personalized products and services that quickly anticipate changing needs and behaviors. There are small and medium-sized financial institutions that can lose customers because of offers that do not favor the customer experience. It is difficult to create analytical models using behavioral data fast enough and precisely to target and predict the products that some customers need in near real time. A similar problem exists in detecting fraud to find patterns of unusual behavior. Financial institutions are estimated to lose between $10 billion and $ 40 billion in revenue annually due to fraud and poor data management practices. For customer targeting and forecast modeling, quantum computing could be a game changer. The data modeling capabilities of quantum computers are expected to be superior in finding models, performing classifications, and making predictions that are not possible today with conventional computers due to the challenges of complex data structures. Another use case in the world of finance is risk analysis. Risk analysis calculations are difficult because it is difficult to analyze many scenarios. Compliance costs are expected to more than double in the coming years. Financial services institutions are under increasing pressure to balance risk, hedge positions more effectively and perform a wider range of stress tests to comply with regulatory requirements. Today, Monte Carlo simulations - the preferred technique for analyzing the impact of risk and uncertainty in financial models - are limited by the scaling of the estimation error. Quantum computers have the potential to sample data differently by testing more results with greater accuracy, providing quadratic acceleration for these types of simulations.

Molecular modeling allows for discoveries such as more efficient lithium batteries. Quantum computing will empower model atomic interaction much more precisely and at much larger scales. We can give again the example of the caffeine molecule. New materials will be able to be used everywhere, whether in consumer products, cars, batteries, etc. Quantum computing will allow molecular orbit calculations to be performed without approximations. Another application is the optimization of a country's electricity network, more predictive environmental modeling and the search for energy sources with lower emissions. Aeronautics will also be a source of use cases. For each landing of an airplane, hundreds of operations are set up: crew change, refueling, cleaning the cabin, baggage delivery, or inspections. Each transaction has sub operations. (The refueling requires a tanker available, a truck driver and two people to fill, in advance it must be sure that the tanker is full). So, in total hundreds of operations and that for only one aircraft landing in limited hours. Now, with hundreds of aircraft landing and sometimes delayed flights, the problem is becoming more and more complex. It is then necessary in real time to recalculate everything for all planes. Electric vehicles have a weakness: the capacity and speed of charging their batteries. A breakthrough in quantum computing made by researchers from IBM and Daimler AG (37), could help meet this challenge. The car manufacturer Daimler is very interested in the impact of quantum computing on the optimization of transport logistics, up to predictions on future materials for electric mobility, in particular the next generation of batteries. There is every reason to hope that quantum computers will yield initial results in the years to come to accurately simulate the chemistry of battery cells, aging processes and the performance limits of battery cells.

      The problem of the commercial traveler can be extended in many fields such as energy, telecommunications, logistics, production chains or resource allocation. For example, in sea freight, there is a great complexity in the management of containers from start to finish: loading, conveying, delivering then unloading in several ports in the world is a multi-parameter problem can be addressed by quantum computing. A better understanding of the interactions between atoms and molecules will make it possible to discover new drugs. Detailed analysis of DNA sequences will help detect cancer earlier by developing models that will determine how diseases develop. The advantage of quantum will be to analyze in detail on a scale never reached the behavior of molecules. Chemical simulations will allow the discovery of new drugs or better predict protein structures, scenario simulations will better predict the risks of a disease or its spread, the resolution of optimization problems will optimize the chains of distribution of drugs, and finally the use of AI will speed up diagnoses, analyze genetic data more precisely.

## Conclusion

The data center of tomorrow is a data center made up of heterogeneous systems, which will run heterogeneous workloads. The systems will be located as close as possible to the data. Heterogeneous systems will be equipped with binary, biological inspired and quantum accelerators. These architectures will be the foundations to address challenges. Like an orchestra conductor, the hybrid cloud will make it possible to set these systems to music thanks to a layer of security and intelligent automation.